\documentclass[letterpaper]{article} 
\usepackage{arxiv}  

\usepackage[final]{microtype}     
\fussy                            
\usepackage{caption}
\usepackage{enumitem}
\setlist{nosep,leftmargin=*,labelsep=0.5em}

\usepackage{times}  
\usepackage{helvet}  

\usepackage{courier}  
\usepackage[hyphens]{url}  
\usepackage{graphicx} 
\usepackage[round,authoryear]{natbib}
\usepackage{caption}
\usepackage{xcolor}

\usepackage{tabularray}
\usepackage{kotex}
\usepackage{colortbl}
\usepackage{amsmath}
\usepackage{booktabs}
\usepackage{arydshln}
\usepackage{boldline}
\usepackage{makecell}
\usepackage{multirow}
\usepackage{subcaption}

\usepackage{algorithm}
\usepackage{algorithmic}

\usepackage{newfloat}
\usepackage{listings}
\DeclareCaptionStyle{ruled}{labelfont=normalfont,labelsep=colon,strut=off}
\floatstyle{ruled}
\newfloat{listing}{tb}{lst}{}
\floatname{listing}{Listing}

\title{Disproportionate Voices: Participation Inequality and Hostile Engagement in News Comments}
\author {
    Sangbeom Kim\textsuperscript{\rm 1}\equalcontrib,
    Seonhye Noh\textsuperscript{\rm 2}\equalcontrib
}
\affiliations {
    \textsuperscript{\rm 1}Department of Economics, Seoul National University\\
    \textsuperscript{\rm 2}Department of Communication, University of California, Los Angeles
}
\usepackage{bibentry}

\begin{document}
\maketitle
\begin{abstract}
Digital platforms were expected to foster broad participation in public discourse, yet online engagement remains highly unequal and underexplored. This study examines the digital participation divide and its link to hostile engagement in news comment sections. Analyzing 260 million comments from 6.2 million users over 13 years on \textit{Naver News}, South Korea’s largest news aggregation platform, we quantify participation inequality using the Gini and Palma indexes and estimate hostility levels with a \textit{KC-Electra} model, which outperformed other Korean pre-trained transformers in multi-label classification tasks. The findings reveal a highly skewed participation structure, with a small number of frequent users dominating discussions, particularly in the Politics and Society domains and popular news stories. Participation inequality spikes during presidential elections, and frequent commenters are significantly more likely to post hostile content, suggesting that online discourse is shaped disproportionately by a highly active and often hostile subset of users. Using individual-level digital trace data, this study provides empirical insights into the behavioral dynamics of online participation inequality and its broader implications for public digital discourse.
\end{abstract}

\section{Introduction}
Digital platforms were once expected to foster broad and equitable participation in public discourse \citep{papacharissi2004democracy}. However, growing evidence suggests that online engagement remains highly unequal, with a small fraction of users dominating digital conversations, potentially skewing public discourse  \citep[e.g.,][]{van_mierlo_1_2014, gasparini_participation_2020, carron-arthur_describing_2014, baqir_beyond_2023, antelmi_characterizing_2019}. The ‘90-9-1’ principle, although not rigorously tested, suggests a significant disparity in online participation, where 90\% of users (’lurkers’) primarily observe without participating, 9\% (’contributors’) engage occasionally, and a mere 1\% (’superusers’) generate the majority of online content \cite{nielsen_90-9-1_2006}.

This study examines the digital participation divide and its relationship with hostile engagement in online news discussions. Using a 13-year dataset from \textit{Naver News}, South Korea’s largest news aggregation platform, we analyze 260 million comments from 6.2 million users to assess the participation inequality between frequent and infrequent commenters in news comment sections and its connection with content hostility. We employ the Gini and Palma indices to quantify participation disparities and apply a transformer-based deep learning model (\textit{KC-Electra}), fine-tuned for multi-label classification, to classify comment hostility levels.

The findings reveal a highly unequal participation structure, with a small number of frequent users contributing disproportionately to news comment sections. This participation divide is particularly pronounced in political news domains and in more widely read news stories. Notably, participation inequality spikes during presidential elections, suggesting that major political events exacerbate engagement disparities. Moreover, these frequent commenters are significantly more likely to post hostile content, including both uncivil and hateful content, suggesting that online discourse is shaped disproportionately by a highly active and often hostile subset of users.

This study makes a novel contribution by systematically linking participation inequality with multiple forms of hostile engagement at scale, using user-level trace data over 13 years. Unlike prior work that typically examines hostility or inequality in isolation, this research uncovers how extreme user imbalances not only dominate the volume of discourse but also disproportionately introduce uncivil and hateful language into public news spaces.

\section{Digital divide and Online Hostility}
Research on digital participation has long documented significant disparities across online platforms. Contrary to early expectations that digital spaces would foster widespread civic participation \citep{papacharissi2004democracy}, the "90-9-1" principle suggests that 90 percent of users passively consume content, 9 percent contribute occasionally, and only 1 percent generate the majority of online content  \cite{nielsen_90-9-1_2006}. Although comprehensive research on this inequality remains scarce, several studies confirm that only a small fraction of users actively participate in digital spaces \citep[e.g.,][]{van_mierlo_1_2014,gasparini_participation_2020,carron-arthur_describing_2014,baqir_beyond_2023,antelmi_characterizing_2019}.

The inequality of digital participation nevertheless remains largely unexplored. Most studies on the digital divide have focused on disparities in physical access to digital systems \cite{chaqfeh_towards_2023} or differences in digital skills and literacy \cite{hargittai2018digital, hargittai2015mind}, with far less attention given to other dimensions of digital inequality \cite{korovkin_towards_2023, scheerder2017determinants, van_dijk_digital_2006}. Thus, there is limited understanding of the extent of participation inequality among individuals who have access to digital platforms but engage with them to varying degrees.

Prior research also suggests that digital participation inequality may be linked to a higher likelihood of hostile engagement. Hostility or incivility in online spaces has been widely documented, particularly in political discussions and news comment sections \citep[e.g., ][]{coe2014online, humprecht_hostile_2020,rowe2015civility, santana2014virtuous,rossini_beyond_2022}. In online comment sections, frequent users are more likely to post hostile content. For example, research on Facebook found that highly engaged users exhibit greater levels of toxicity in their comments \cite{kim2021distorting}. Similarly, studies on news comment sections indicate that hostility tends to cluster among the most active participants \cite{humprecht_hostile_2020,rowe2015civility}, potentially shaping broader public perceptions of digital discourse.

The potential association between frequent commenting and hostile content may be driven by anger, a high-arousal emotion that is strongly linked to greater engagement and participation \cite{berger2011arousal,brady2017emotion,crockett2017moral,hasell2016partisan,masullo2021does,valentino2011election}. This pattern is particularly pronounced in partisan digital environments, where hostility toward out-groups generates higher engagement than in-group favoritism \cite{rathje2021out,yu2024partisanship}. \citeauthor{masullo2021does} \shortcite{masullo2021does} further suggests that anger increases the likelihood of users actively expressing their opinions online, regardless of the opinion climate they encounter.

Building on these insights, this study advances research on the digital divide by bridging two critical aspects of online engagement—digital participation inequality and online hostility—that have not been systematically examined together. By leveraging individual-level news comment behavior data over a 13-year period, this study provides a rare opportunity to examine both the severity of the participation divide between frequent and infrequent users and whether this divide is indeed linked to hostile engagement.
\section{Data}
\subsection{\textit{Naver News}}
South Korea is one of the most digitally connected countries in the world, boasting the highest percentage of high-speed broadband connections among OECD nations \cite{pak_digitalization_2021}. In addition, in this country, online news consumption is overwhelmingly concentrated on news aggregator platforms rather than individual news websites. According to a global comparison of 46 countries, South Korea had the highest rate of news consumption via news aggregators and the lowest rate via direct access to news websites in 2021 \cite{oh2021digital}. Among these platforms, \textit{Naver News} stands as the most dominant, reflecting its unparalleled role in shaping the country’s digital news ecosystem. Over 90 percent of Koreans use \textit{Naver} as their primary search engine, and 87 percent rely on \textit{Naver News} for their online news consumption \cite{kpf_media_2021}. This shows that the inequality of digital access is at least very little at play.

This minimal digital access inequality ensures that disparities in online engagement are not driven by differences in basic access to digital infrastructure but rather by individual preferences and behavioral choices. Unlike in countries where digital divides are primarily shaped by disparities in internet access, South Korea presents a unique context where virtually all users have the opportunity to engage with news content online, allowing for a more precise examination of participation inequality in digital discourse.

The platform, \textit{Naver News} offers users free access to news content from major news outlets in the country. A key feature of the platform is its in-link system, which enables users to read full articles and comment on them directly \textit{within} the platform, rather than being redirected to the original news websites. This design eliminates the need for users to create accounts on multiple media sites, effectively centralizing news consumption and discussion within a single platform.

The comprehensive scope of \textit{Naver News} and its centralized commenting system make its data particularly valuable for studying digital participation and hostile engagement at the individual level. Because South Korea has minimal barriers to internet access, participation disparities on the platform likely reflect user preferences rather than structural access limitations. Moreover, \textit{Naver News} data allows for tracking individual commenting behavior over time, providing a rare opportunity to examine participation patterns based on frequency of use.
\subsection{News Comment Data}
From Naver News, we collected approximately 260 million comments along with unique user identifiers (different from actual accounts; partially masked by the platform) from January 2008 to September 2020, using the R package \textit{N2H4}. During this period, Naver News published a daily list of the 30 most-read articles across six domains—Politics, Society, Economy, World, IT/Science, and Life/Culture—amounting to 180 articles per day. The final dataset contains 802,946 articles from 141 outlets and 260,203,552 comments posted by approximately 6,170,121 unique users. On average, each article received 324 comments. The volume of commenting activity increased significantly over time (see \textit{Appendix}), likely reflecting the growing accessibility of online news platforms.

Given the large volume of comments and the presence of heavily active users, we conducted a supplementary analysis to assess whether certain users exhibited non-authentic(bot-like) behavior, such as repetitive commenting. We find that while a small subset of users show highly duplicated(or similar) content, their overall prevalence is limited and unlikely to bias the main results(see \textit{Appendix}).
\begin{figure}[H]
    \centering
    \includegraphics[width=0.9\columnwidth]{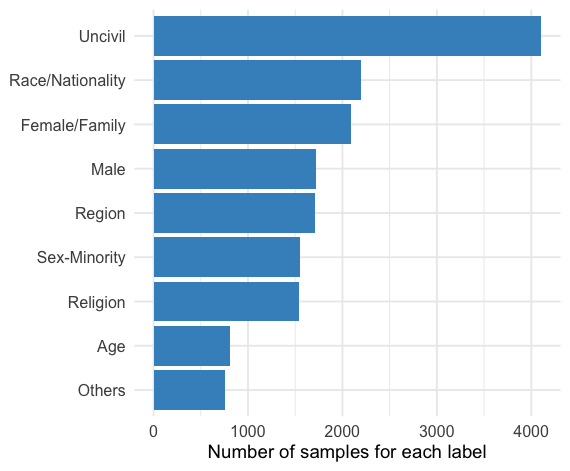}
    \caption{Distribution of Hateful and Uncivil Sentences. 'Civil' sentences are excluded in this figure. We allow for overlapping counts here. If a sentence has two labels, it will be counted once for each label.}
    \label{fig1}
\end{figure}
\subsection*{Hate Speech Data}
Previous research distinguishes between intolerant messages, which express harmful or discriminatory intent toward specific groups, and impolite posts, which contain rude or offensive language \cite{rossini_beyond_2022}. Although terminology varies across the literature \citep[e.g.,][]{rossini_beyond_2022, rowe2015civility, rega2023incivility}, these studies emphasize that not all uncivil messages are equally damaging to democratic discourse, underscoring the need to differentiate between the two types.
This study adopts a typology of “hostility” that distinguishes these forms: we refer to generally rude or offensive content as \textit{uncivil}, and content targeting specific social groups as \textit{hateful}.

To detect hostility in news comments while capturing this distinction, we trained and compared multiple transformer-based models -including BERT and Electra- on the \textit{Korean Unsmile Dataset} \cite{SmilegateAI2022KoreanUnSmileDataset}. 
This dataset includes ten distinct labels (civil, uncivil, and hate speech targeting nine different social groups), allowing us to evaluate model performance across nuanced types of hostile languages. The ten categories specifically include \textit{Civil} (devoid of hate speech), \textit{Uncivil} (disparaging language or personal attacks), and various hate speech types targeting \textit{race/nationality}, \textit{region}, \textit{gender}(male/female/family), \textit{religion}, \textit{age}, and \textit{sexual minorities}. Each comment can receive multiple labels across categories.

However, this dataset may potentially misclassify neutral comments as hateful \citep{kang2022korean}. For example, a benign statement referencing a group may be flagged as hate speech incorrectly. We supplemented the dataset with additional neutral sentences following \citeauthor{kang2022korean} \shortcite{kang2022korean} to mitigate this issue. In the training dataset, uncivil content is the most frequent category (24.5\%), followed by hateful content targeting race/nationality (13\%), female/family (12\%), male (11\%), region (10\%), religion (9\%), sex minority (9\%), and age (4.8\%). These frequencies are shown in Figure \ref{fig1}.

While Figure \ref{fig1} shows label-wise frequencies, it does not reflect how many comments carry multiple labels. To better capture this multi-label structure, we provide further descriptive analysis of label co-occurrence patterns and overlap among hate categories in \textit{Appendix}.

\section{Methods}
\subsection*{Measuring Participation Inequality}

To assess user engagement levels, we first ranked all users in the dataset based on the number of comments they posted, with the most active commenters placed at the top. This ranking allowed us to classify users into different engagement groups, which were then used to compare hostility levels in their comments. Our analysis primarily focuses on the top 10\% of the most active commenters, comparing them to the bottom 40\% of commenters, who exhibit significantly lower engagement.

To quantify participation inequality among these user groups, we employed two widely used economic disparity metrics: the Gini index and the Palma index \cite{atkinson1970measurement, kakwani1977applications}, both of which have been applied in prior research to assess engagement inequalities in digital spaces  \cite{shu_user_2020}.

The Gini index measures the overall dispersion of participation levels, reflecting how unequally comments are distributed among users. A higher Gini index indicates greater inequality in engagement. However, the Gini index has notable limitations in interpretation. Two distributions with identical Gini values can have different underlying structures, making it difficult to capture whether disparities are driven by the most or least active users. Additionally, the Gini index is more sensitive to changes in the middle of the distribution but less responsive to variations at the top and bottom.

To address these limitations, we incorporate the Palma index, which specifically measures the ratio of participation between the top 10\% of commenters and the bottom 40\%. An increasing Palma index indicates that the most active users are gaining even greater dominance over the least active users, highlighting the skewed nature of digital participation. Unlike the Gini index, the Palma index provides a clearer interpretation of who dominates the discourse in digital spaces and to what extent.

We applied these two metrics across different time periods, news domains, and news popularity rankings, depending on the specific analytical focus of each part of the study.

\subsection*{Measuring Contribution to Inequality}
After calculating the inequality metrics, we assess whether the observed disparities are primarily driven by frequent or infrequent commenters using the relative mean deviation (RMD). This metric is mathematically defined as follows:
\begin{equation}
RMD_{ig} = \frac{N_i-\mu_g}{\mu_g}    
\end{equation}
where $i$ represents an individual user, $g$ denotes the news domain. $N_i$ is the number of comments posted by user $i$, and $\mu_g$ represents the average number of comments per user in news domain $g$.

The RMD serves as a counterfactual measure to evaluate participation inequality. In a scenario where all users contributed an equal number of comments, the comment space would exhibit perfectly equal participation. This hypothetical equal participation level is represented by $\mu_g$. By comparing each user's actual comment count to $\mu_g$, the RMD quantifies how much more or less each user contributes relative to this counterfactual equality.

This metric allows us to determine whether inequality is driven by frequent commenters posting significantly more than expected or by infrequent commenters contributing far less than the counterfactual amount. In doing so, it provides a clearer picture of how participation disparities emerge in online discussions.

\subsection*{Measuring Comment Hostility}
\subsubsection{Basic Framework}
To assess the level of hostility in user comments, we conducted a content analysis by stratifying commenters into heavy (top 10\%) and light (bottom 40\%) engagement groups, based on the Palma index. Within the top 10\%, we further isolated the top 1\% of commenters, as a small subset appeared disproportionately frequently compared to others. 

As an initial step, we \textit{KC-BERT} (Base and Large) and \textit{KC-Electra} \cite{lee2020kcbert} models using the hate speech dataset described earlier. These KC-specific models are designed to better capture the nuances of Korean online comments, including informal variations and synonymous expressions. To benchmark their performance and assess classification robustness, we also trained two widely used reference models: \textit{KoBERT} and \textit{KoElectra}.

After model training, we selected the best-performing model and applied it to a 1\% stratified sample of comments from each user group. The model assigned multi-label scores to each comment, and for simplicity, we retained only the highest-scoring label per comment, filtering out those for which all scores were below 0.5. We then consolidated the ten original hate categories into three broader classes: \textit{civil}, \textit{uncivil}, and \textit{hateful}. This grouping allows for a more precise comparison of hostility level across commenter groups by reducing label complexity while preserving key semantic distinctions. 

Specifically, \textit{uncivil} comments include general profanity and personal attacks, whereas \textit{hateful} comments contain derogatory or discriminatory expressions targeted at specific social groups (e.g., gender, religion, race, or region). Comments lacking such content are labeled \textit{civil}. Based on this three-way classification, we then compared the distribution of component types (\textit{civil}, \textit{uncivil}, and \textit{hateful}) across user engagement groups. Using a chi-squared test of proportions, we tested whether the observed differences in hostility levels between heavy and light commenters were statistically significant. 

\subsubsection{Details of the Training Process}
To determine optimal performance, we conducted experiments across a range of hyperparameters: learning rates of 2e-5, 3e-5, and 5e-5; batch sizes of 8, 16, 32, and 64; and training epochs of 3 and 5. All models were trained and evaluated using a single NVIDIA Tesla T4 GPU on Google Colab. The training for each model took approximately 2–4 hours, depending on model size and batch configuration. 

Table 1 summarizes the best-performing configurations. Among all models, \textit{KC-Electra} slightly outperformed the others. This result aligns with our expectations, as KC-specific models are pretrained on Korean corpora that include substantial amounts of informal, user-generated content, such as online comments, making them more attuned to the linguistic characteristics of our dataset.

Model performance was evaluated primarily using the Label Ranking Average Precision (LRAP) score. Additional metrics—including precision, recall, and F1 Score—also confirmed the superior performance of KC-based models in handling the multi-label classification of hostile and hate speech in Korean text.

\begin{table}[H]
\centering
\begin{tabular}{lcccc}
\toprule
\textbf{Model} & \textbf{F1} & \textbf{Precision} & \textbf{Recall} & \textbf{LRAP} \\
\midrule
KC-BERT Base   & 0.862 & 0.876 & 0.850 & 0.926 \\
KC-BERT Large  & 0.865 & \textbf{0.884} & 0.851 & 0.928 \\
KC-Electra     & \textbf{0.872} & 0.882 & \textbf{0.866} & \textbf{0.931} \\
KoBERT         & 0.839 & 0.837 & 0.843 & 0.908 \\
KoElectra      & 0.849 & 0.847 & 0.852 & 0.913 \\
\bottomrule
\end{tabular}
\caption{Performance comparison of transformer-based models on the multi-label hate speech classification task. The best-performing model for each metric is shown in bold.}
\label{tab:model-performance}
\end{table}

\section{Participation Inequality}
Descriptive statistics on participation levels indicate a stark digital participation gap (Figure 3). On average, the top 10\% of frequent commenters account for nearly half of all comments in news comment sections (50.11\%), while the least active half (bottom 50\%) contributes only 14.99\% of total comments over the years. The figure clearly illustrates a consistent and substantial divide in digital participation, where a small subset of users disproportionately dominates the conversation. This imbalance underscores the motivation for our study, highlighting the need to investigate the structural disparities in online engagement.
\begin{figure}[ht]
    \centering
    \includegraphics[width=0.9\columnwidth]{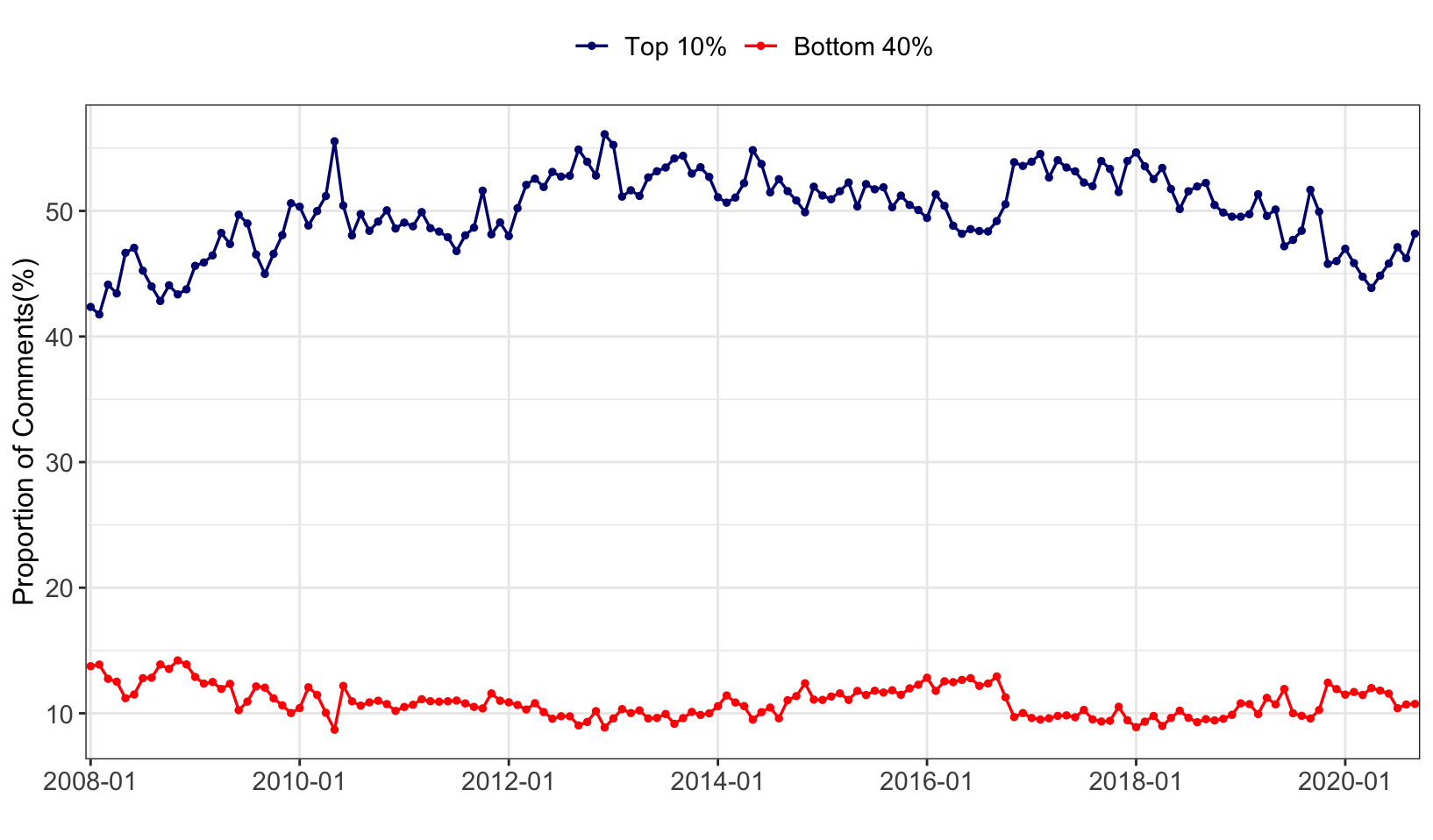}
    \caption{Share of Comments by Top 10\% and Bottom 40\% Gropus}
    \label{fig2}
\end{figure}

\subsection*{Participation Inequality by News Domain and Popularity}
To further examine this divide, we quantified participation inequality within the news ecosystem using the Gini index and the Palma index. We then compared participation inequality (a) across six news domains (\textit{Politics, Society, Economy, World, IT/Science, and Life/Culture}) and (b) at varying levels of news popularity. Note that \textit{Naver News} publishes a daily list of the 30 most-read articles, referred to as ‘\textit{Ranking News}.’ To measure news popularity, we used these rankings, with 1st representing the least popular and 30th the most popular article of the day. We then calculated Gini and Palma indexes for different news stories based on their popularity ranks to assess how inequality changes across news interest levels.

\begin{figure}[ht]
\centering
\includegraphics[width=0.85\columnwidth]{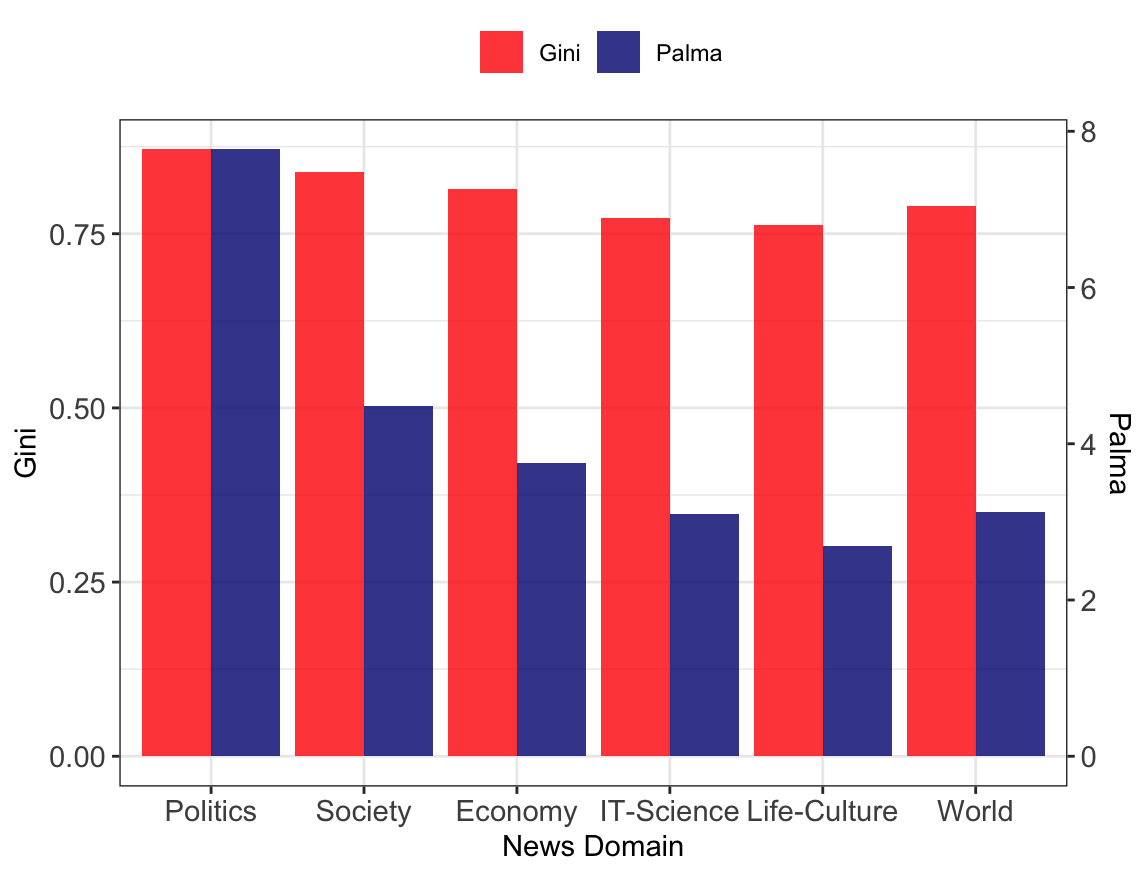} 
\caption{The Gini and Palma index Over Time by News Domain}
\label{fig3}
\end{figure}

Figure \ref{fig3} illustrates participation inequality across different news domains, showing that political news exhibits the highest levels of inequality compared to other categories. Both the Gini and Palma indices reveal that Politics consistently stands out as the most unequal domain, indicating that discussions in political news sections are dominated by a small subset of highly active commenters. Society and Economy also exhibit relatively high participation inequality, though to a lesser extent than Politics. In contrast, domains such as Life/Culture and IT/Science display lower levels of inequality, suggesting that discussions in these categories are more evenly distributed among users. 

\begin{figure}[ht]
\centering
\includegraphics[width=0.9\columnwidth]{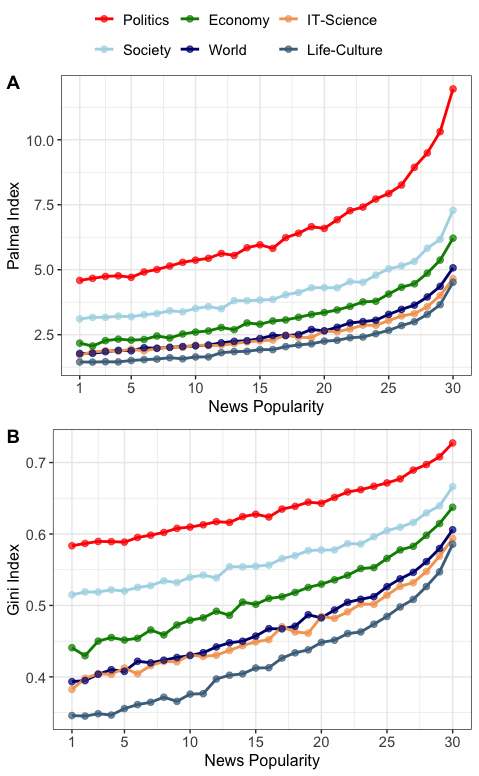} 
\caption{The Palma(Panel A) and Gini(Panel B) index by News Popularity}
\label{fig4}
\end{figure}

Figure \ref{fig4} presents participation inequality as measured by the Palma index (Panel A) and the Gini index (Panel B) across different levels of news popularity. Across all domains, both indices show a clear upward trend, indicating that as a news story becomes more popular, participation inequality increases. This pattern suggests that highly popular articles tend to be dominated by a small group of frequent commenters, while less popular articles see a more balanced distribution of participation. Among the different news domains, Politics and Society, again, consistently exhibit the highest levels of inequality across all levels of popularity, reinforcing the idea that digital participation gaps are most pronounced in politically charged discussions.

Taken together, these findings suggest that participation inequality is not only domain-specific but also influenced by news popularity. The more widely read an article is, the more concentrated the conversation becomes among a small subset of highly active users, particularly in Politics and Society.

\begin{figure*}[ht]
\centering
\includegraphics[width=0.95\textwidth]{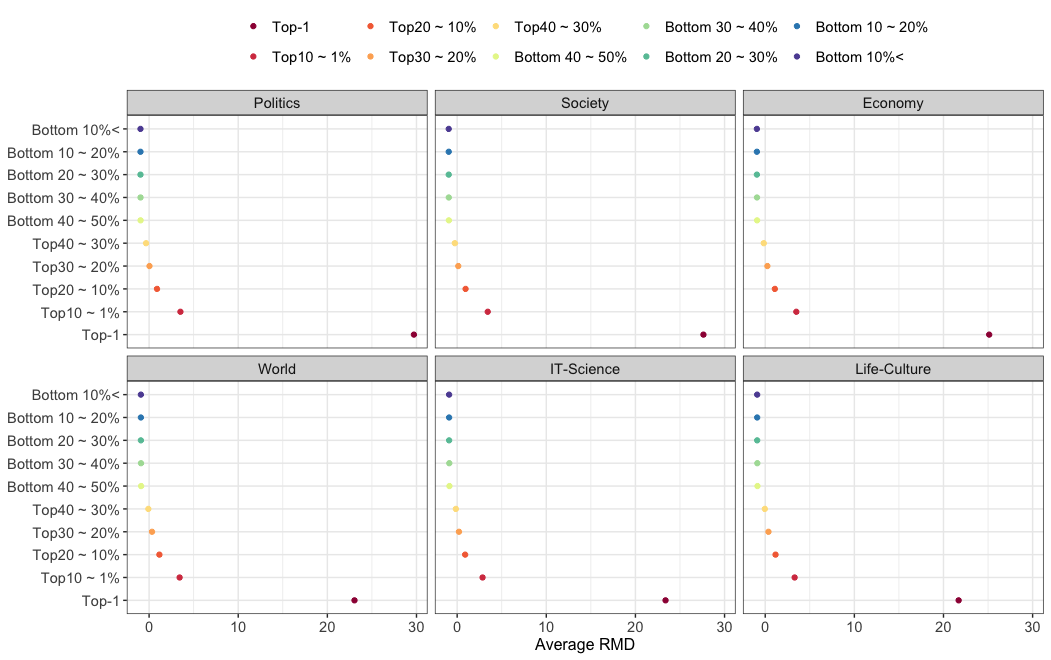} 
\caption{Average Relative Mean Deviation by News Domain}
\label{fig5}
\end{figure*}
\subsection*{User Contribution to Participation Inequality}
To assess which user groups contribute most to participation inequality, we analyzed Relative Mean Deviation (RMD) scores. While the Palma and Gini indices measure overall inequality, they do not reveal how different user groups contribute to these disparities. RMD addresses this gap by indicating how much each group's participation deviates from a hypothetical benchmark of perfect equality, where all users contribute an equal number of comments within a given news domain and news popularity level. A value of 0 represents perfect equality, while negative values indicate lower-than-expected participation, and positive values indicate excessive participation relative to the equality benchmark.

Figure \ref{fig5} presents RMD scores across different user groups, segmented into ten participation levels to capture finer distinctions beyond the broad bottom 40\% and top 10\% classifications. The figure shows that the least active commenter groups (Bottom 10\% to Top 30-20\%) cluster around zero, indicating that their participation closely aligns with the expected equal participation benchmark. 

In contrast, there is a progressive and disproportionate increase in deviation among more active users, with the top 1\% of commenters exhibiting the highest deviation. The top 1\% of users have an RMD between 23 and 30, compared to an average deviation of 3 among other active groups, demonstrating their outsized influence on digital discourse.

These findings underscore two key aspects of participation inequality. First, they indicate that the observed participation gap is primarily driven by highly active users posting disproportionately more comments, rather than infrequent users posting significantly fewer comments. This suggests that participation inequality is a function of over-contribution by a small subset of users rather than disengagement by the majority. 

Second, there is a sharp divide even among active commenters, particularly between the top 1\% and the rest, highlighting that the most extreme contributors play a dominant role in shaping discussions. This suggests that online discourse is not only concentrated among a small subset of users but is further skewed by an even smaller group of hyperactive commenters, reinforcing the severe imbalances in digital participation.

\subsection*{Participation Inequality and Political Events}
Beyond these structural patterns, we now examine how participation inequality fluctuates in response to major political events, particularly during South Korea’s electoral cycles (2012 and 2017 presidential elections) and one of the most significant political events of the study period—the 2016 impeachment of President Park Geun-hye. 
\begin{figure}[ht]
\centering
\includegraphics[width=0.8\columnwidth]{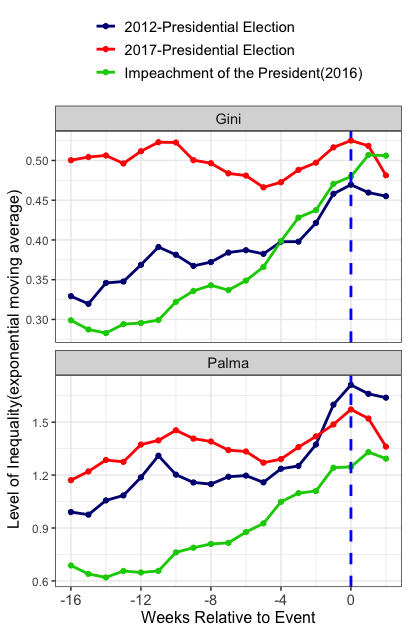} 
\caption{Participation Inequality Leading Up to Presidential Elections and the 2017 Impeachment}
\label{fig6}
\end{figure}
Figure \ref{fig6} illustrates the Gini and Palma indices in the weeks leading up to three key political events. The trends suggest that participation inequality intensifies as major political events approach, with both indices showing a marked increase in the final weeks leading up to each event. A small subset of highly active users becomes even more dominant in news comment sections during politically charged periods, further exacerbating the imbalance in online discourse. 

While the Gini coefficient does not appear to increase further during the 2017 presidential election, this should not be interpreted as an absence of change. Rather, the spike in participation inequality observed following the 2016 impeachment appears to have persisted into the 2017 election period. These findings suggest that political events act as catalysts for deepening participation inequality, amplifying the influence of highly engaged users while sidelining less active participants.

\section{Comment Hostility}
\begin{figure}[ht]
\centering
\includegraphics[width=0.8\columnwidth]{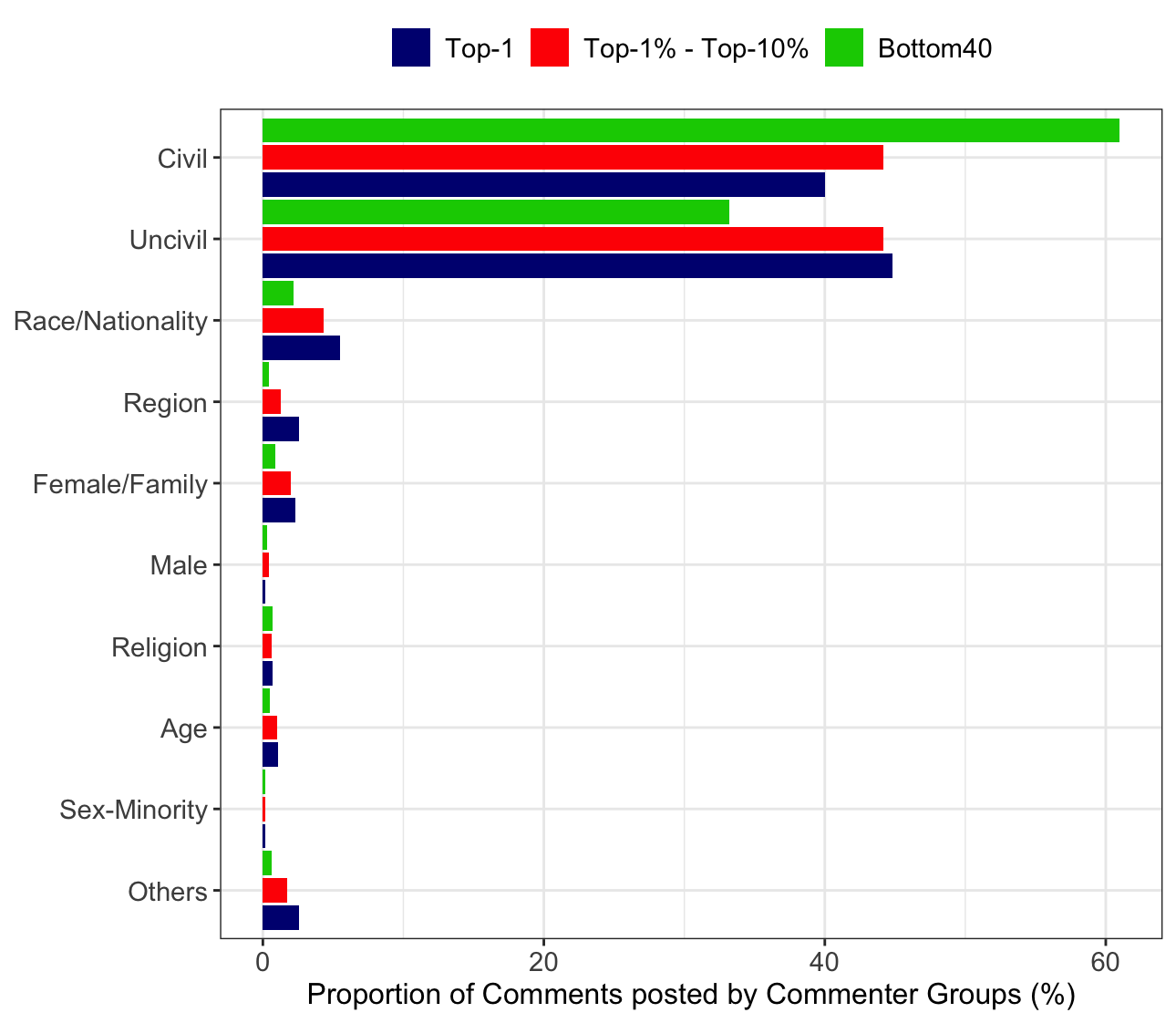} 
\caption{Hate Comment Classification Result by Percentile User Group}
\label{fig7}
\end{figure}
\begin{figure}[ht]
\centering
\includegraphics[width=0.8\columnwidth]{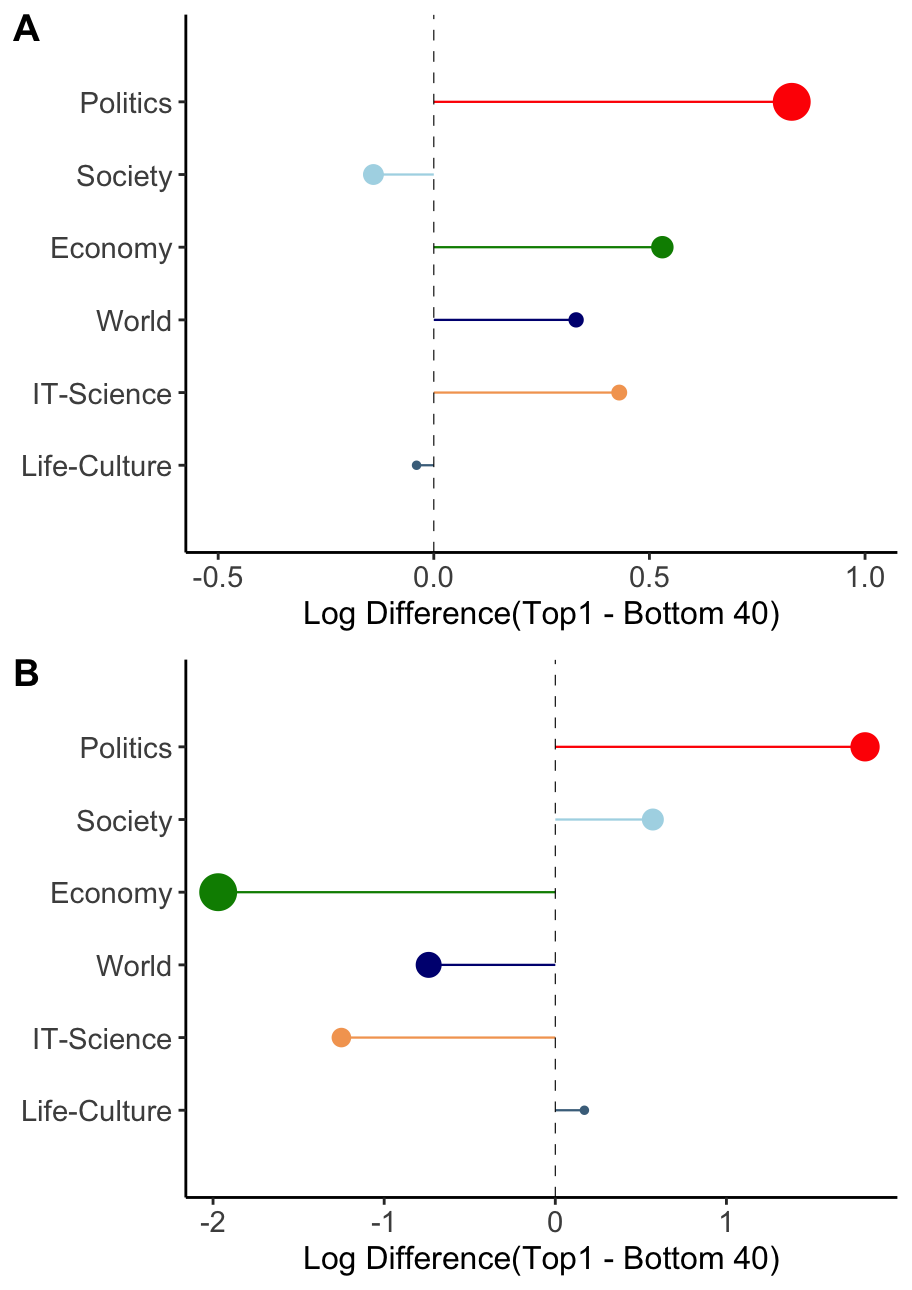} 
\caption{Log Difference in the Proportion of Uncivil (Panel A) and Hateful Comments (Panel B) between Extreme (Top 1\%) and Inactive (Bottom 40\%) User Group Across News Domains. Point sizes indicate the absolute difference in proportion.}
\label{fig8}
\end{figure}
Previous studies suggest that more active users in comment sections are more likely to exhibit hostility. To examine this, we conducted a computational content analysis to assess the levels of hostility in comments posted by different user groups. For this analysis, we focused on three distinct commenter groups, ranked by their commenting activity: (1) the top 1\% most active commenters, (2) the next most active group (top 1–10\% (excluding top 1\%)), and (3) the bottom 40\% least active commenters. It is important to note that the top 1\% and top 1–10\% are distinct groups, unlike the broader categories used in prior analyses. 

Given the unique behavior of the most active users, as shown in the participation inequality results, we isolated the top 1\% separately to better capture the extreme engagement patterns of this highly active subset. For each group, we randomly selected 1\% of comments from the raw dataset for analysis. These comments were then classified as either (1) civil, (2) uncivil, or (3) one of eight types of hateful comments using a deep learning classifier trained on a large dataset of labeled comments. 

Figure \ref{fig7} presents the distribution of comment categories across these three user groups. For simplicity and clarity, we aggregated the original ten fine-grained labels into three macro‐categories—\textit{civil}, \textit{uncivil}, and \textit{hateful}—and report these simplified proportions in Table \ref{tab:comment_proportion}, facilitating a concise comparison of hostility across groups.

As expected, the most frequent commenters—the top 1\% and top 10\%(excluding the top 1\%)—are significantly more likely to post \textit{uncivil} comments compared to the less active bottom 40\%, as confirmed by chi-square proportion tests (p $<$ 0.001 for the comparison between bottom 40\% and top 1–10\%, and between bottom 40\% and top 1\%). 

To complement this finding, we computed Cohen's $h$ \citep{cohen1988statistical}, which indicated that these differences are not only statistically significant but also non-trivial in magnitude. All comparisons yielded h $\geq$ 0.2, which meets the threshold for a small effect size (see \textit{Appendix} for interpretation criteria), with the largest effect in the civil category (h = 0.34 for bottom 40 \% vs. top 1–10\%; h = 0.42 for bottom 40 \% vs. top 1 \%), confirming that the observed differences—though driven primarily by shifts in civility rather than target‑specific hate—are substantively meaningful rather than mere artifacts of large sample size. Together with the chi‑square results, these findings provide strong evidence of behavioral divergence in commenting style across user activity levels. 

Regarding \textit{hateful} content, the divide in online hostility extends even among active users: the top 1\% is significantly more likely to post hateful comments than the top 1–10\% (chi-square test, p $<$ 0.001). This finding further reinforces the digital participation divide, showing that not only do a small number of users dominate discussions, but they also tend to engage in higher levels of incivility and hate speech.
\begin{table}[ht]
\centering
\renewcommand{\arraystretch}{1.2}
\begin{tabular}{lccc}
\toprule
\makecell{\textbf{User}\\\textbf{Group}} & \textbf{Civil} & \textbf{Uncivil} & \textbf{Hate} \\
\midrule
Bottom 40\% & 0.610 & 0.332 & 0.058 \\
Top 1\% - Top 10\% & 0.442 & 0.442 & 0.116 \\
Top 1\% & 0.400 & 0.448 & 0.152 \\
\bottomrule
\end{tabular}
\caption{Proportion of comment categories by user group}
\label{tab:comment_proportion}
\end{table}
\begin{table}[H]
\centering
\begin{tabular}{@{}lccc@{}}
\toprule
\textbf{Comparison} & \textbf{Civil} & \textbf{Uncivil} & \textbf{Hate} \\
\midrule
Bottom 40 vs Top 1\% - Top 10\% & 0.34 & 0.23 & 0.21 \\
Bottom 40 vs Top 1\%         & 0.42 & 0.24 & 0.31 \\
\bottomrule
\end{tabular}
\caption{Effect sizes (Cohen's $h$) across label categories by group comparison}
\label{table;cohens}
\end{table}

We replicated the analysis using four alternative Korean language models to ensure that our classification results are not overly dependent on model choice. The results consistently show that more active users tend to post a higher proportion of hate comments across all models. The full comparison of raw proportions and effect sizes (Cohen’s $h$) is provided in \textit{Appendix}

The disparity in hostility between active and inactive groups is still evident when examining differences across news domains. As shown in Figure \ref{fig8}, the gaps in both uncivil and hateful comment proportions are particularly pronounced in the Politics domain, suggesting that highly engaged users are especially likely to contribute hostile discourse in political discussions.
\section{Conclusion}
This study underscores the stark participation inequality in online news comment sections, where a small but highly active subset of users disproportionately shapes digital discourse. Analyzing 260 million comments over 13 years on \textit{Naver News}, we find that this participation gap is particularly pronounced in political news discussions and highly popular news stories, intensifying during major political events such as presidential elections. The analysis also reveals that the most active commenters contribute disproportionately to the overall volume of engagement, further amplifying their influence. Moreover, these frequent commenters are significantly more likely to engage in hostile discourse, posting both uncivil and hateful content at higher rates than less active users. This suggests that online discussions are not only dominated by a small fraction of users but are also skewed toward a more hostile or hateful discourse. 

These findings carry important implications for digital public discourse and online platform governance. The dominance of a small, often hostile group in comment sections raises concerns about the representativeness of online discussions and their potential to skew public perceptions. Platforms aiming to foster healthier discourse may need to consider interventions that encourage broader participation while mitigating the outsized influence of highly engaged yet hostile users. Future research should further explore the causal mechanisms behind these dynamics and investigate potential strategies to counteract digital participation disparities and online hostility.

\section{Limitations}

While this study provides valuable insights into digital participation inequality and hostile discourse, it has several limitations that should be addressed in future research.

First, although our findings reveal a significant disparity in hostility between active and inactive user groups, further analysis is needed to understand the underlying linguistic mechanisms driving this disparity. Specifically, a more granular examination of how hostile language is constructed and varies between these groups would provide deeper insights. However, this presents a methodological challenge due to the complex structure of the Korean language. Korean allows for the creation of new words through character combinations, often leading to non-standard lexical variations in online discussions. This makes tokenization particularly difficult, as conventional NLP methods may fail to capture these variations accurately. Additionally, detecting hostility—especially hateful content targeting specific sociopolitical groups—is further complicated by implicit and coded expressions that may not contain overt hate speech terms but still convey derogatory or exclusionary meanings. This linguistic flexibility enables users to mask hostility, making deep-learning-based classification models prone to under-detection of such content. Addressing this issue requires more sophisticated linguistic processing techniques, such as context-aware tokenization models, morphological analysis tailored to Korean online discourse, and adversarial training methods that can better capture implicit hostility. Future research should refine these approaches to improve the precision of hostility detection, particularly for nuanced forms of incivility and hate speech.

Second, our study does not establish a direct causal relationship between participation inequality and online hostility. While our findings suggest that hostility is more prevalent among highly active users, we have not explicitly tested whether increasing inequality drives greater hostility or if other factors mediate this relationship. As participation inequality intensifies—especially during politically charged periods—aggressive discourse may become more concentrated among dominant commenters. However, our dataset is limited to observational digital trace data, which primarily captures user behaviors, comment timing, and content, but does not account for underlying psychological or social motivations. Future research should explore experimental methods to better understand the causal links between participation inequality and online hostility.

Despite these limitations, this study provides a foundational analysis of how a small proportion of users shapes digital discourse through both disproportionate engagement and elevated hostility. The findings are particularly novel given the scale and granularity of the dataset, as well as the platform’s minimal algorithmic bias—specifically, the absence of personalized ranking in the daily most-read articles—which allows for clearer attribution of behavioral patterns. This advances our understanding of how participation inequality can distort democratic discourse, even in relatively open digital environments.

Addressing these challenges in future research will be crucial for developing more effective moderation strategies and fostering healthier online discussions.

\bibliographystyle{arxiv}
\bibliography{arxiv}

\begin{thebibliography}{39}
\providecommand{\natexlab}[1]{#1}

\bibitem[{Antelmi, Malandrino, and Scarano(2019)}]{antelmi_characterizing_2019}
Antelmi, A.; Malandrino, D.; and Scarano, V. 2019.
\newblock Characterizing the {Behavioral} {Evolution} of {Twitter} {Users} and {The} {Truth} {Behind} the 90-9-1 {Rule}.
\newblock In \emph{Companion {Proceedings} of {The} 2019 {World} {Wide} {Web} {Conference}}, 1035--1038. San Francisco USA: ACM.
\newblock ISBN 978-1-4503-6675-5.

\bibitem[{Atkinson et~al.(1970)}]{atkinson1970measurement}
Atkinson, A.~B.; et~al. 1970.
\newblock On the measurement of inequality.
\newblock \emph{Journal of economic theory}, 2(3): 244--263.

\bibitem[{Baqir et~al.(2023)Baqir, Chen, Diaz-Diaz, Kiyak, Louf, Morini, Pansanella, Torricelli, and Galeazzi}]{baqir_beyond_2023}
Baqir, A.; Chen, Y.; Diaz-Diaz, F.; Kiyak, S.; Louf, T.; Morini, V.; Pansanella, V.; Torricelli, M.; and Galeazzi, A. 2023.
\newblock Beyond {Active} {Engagement}: {The} {Significance} of {Lurkers} in a {Polarized} {Twitter} {Debate}.
\newblock ArXiv:2306.17538 [physics], arXiv:2306.17538.

\bibitem[{Berger(2011)}]{berger2011arousal}
Berger, J. 2011.
\newblock Arousal increases social transmission of information.
\newblock \emph{Psychological science}, 22(7): 891--893.

\bibitem[{Brady et~al.(2017)Brady, Wills, Jost, Tucker, and Van~Bavel}]{brady2017emotion}
Brady, W.~J.; Wills, J.~A.; Jost, J.~T.; Tucker, J.~A.; and Van~Bavel, J.~J. 2017.
\newblock Emotion shapes the diffusion of moralized content in social networks.
\newblock \emph{Proceedings of the National Academy of Sciences}, 114(28): 7313--7318.

\bibitem[{Carron-Arthur, Cunningham, and Griffiths(2014)}]{carron-arthur_describing_2014}
Carron-Arthur, B.; Cunningham, J.~A.; and Griffiths, K.~M. 2014.
\newblock Describing the distribution of engagement in an {Internet} support group by post frequency: {A} comparison of the 90-9-1 {Principle} and {Zipf}'s {Law}.
\newblock \emph{Internet Interventions}, 1(4): 165--168.
\newblock Publisher: Elsevier.

\bibitem[{Chaqfeh et~al.(2023)Chaqfeh, Asim, AlShebli, Zaffar, Rahwan, and Zaki}]{chaqfeh_towards_2023}
Chaqfeh, M.; Asim, R.; AlShebli, B.; Zaffar, M.~F.; Rahwan, T.; and Zaki, Y. 2023.
\newblock Towards a {World} {Wide} {Web} without digital inequality.
\newblock \emph{Proceedings of the National Academy of Sciences}, 120(3): e2212649120.

\bibitem[{Coe, Kenski, and Rains(2014)}]{coe2014online}
Coe, K.; Kenski, K.; and Rains, S.~A. 2014.
\newblock Online and uncivil? Patterns and determinants of incivility in newspaper website comments.
\newblock \emph{Journal of communication}, 64(4): 658--679.

\bibitem[{Cohen(1988)}]{cohen1988statistical}
Cohen, J. 1988.
\newblock \emph{Statistical power analysis for the behavioral sciences}.
\newblock Hillsdale, NJ: Lawrence Erlbaum Associates, 2nd edition.

\bibitem[{Crockett(2017)}]{crockett2017moral}
Crockett, M.~J. 2017.
\newblock Moral outrage in the digital age.
\newblock \emph{Nature human behaviour}, 1(11): 769--771.

\bibitem[{Gasparini et~al.(2020)Gasparini, Clarisó, Brambilla, and Cabot}]{gasparini_participation_2020}
Gasparini, M.; Clarisó, R.; Brambilla, M.; and Cabot, J. 2020.
\newblock Participation {Inequality} and the 90-9-1 {Principle} in {Open} {Source}.
\newblock In \emph{Proceedings of the 16th {International} {Symposium} on {Open} {Collaboration}}, 1--7. Virtual conference Spain: ACM.
\newblock ISBN 978-1-4503-8779-8.

\bibitem[{Glenski, Volkova, and Kumar(2020)}]{shu_user_2020}
Glenski, M.; Volkova, S.; and Kumar, S. 2020.
\newblock User {Engagement} with {Digital} {Deception}.
\newblock In Shu, K.; Wang, S.; Lee, D.; and Liu, H., eds., \emph{Disinformation, {Misinformation}, and {Fake} {News} in {Social} {Media}}, 39--61. Cham: Springer International Publishing.
\newblock ISBN 978-3-030-42698-9 978-3-030-42699-6.
\newblock Series Title: Lecture Notes in Social Networks.

\bibitem[{Hargittai(2018)}]{hargittai2018digital}
Hargittai, E. 2018.
\newblock The digital reproduction of inequality.
\newblock In \emph{The inequality reader}, 660--670. Routledge.

\bibitem[{Hargittai and Shaw(2015)}]{hargittai2015mind}
Hargittai, E.; and Shaw, A. 2015.
\newblock Mind the skills gap: the role of Internet know-how and gender in differentiated contributions to Wikipedia.
\newblock \emph{Information, communication \& society}, 18(4): 424--442.

\bibitem[{Hasell and Weeks(2016)}]{hasell2016partisan}
Hasell, A.; and Weeks, B.~E. 2016.
\newblock Partisan provocation: The role of partisan news use and emotional responses in political information sharing in social media.
\newblock \emph{Human Communication Research}, 42(4): 641--661.

\bibitem[{Humprecht, Hellmueller, and Lischka(2020)}]{humprecht_hostile_2020}
Humprecht, E.; Hellmueller, L.; and Lischka, J.~A. 2020.
\newblock Hostile {Emotions} in {News} {Comments}: {A} {Cross}-{National} {Analysis} of {Facebook} {Discussions}.
\newblock \emph{Social Media + Society}, 6(1): 205630512091248.

\bibitem[{Kakwani(1977)}]{kakwani1977applications}
Kakwani, N.~C. 1977.
\newblock Applications of Lorenz curves in economic analysis.
\newblock \emph{Econometrica: Journal of the Econometric Society}, 719--727.

\bibitem[{Kang et~al.(2022)Kang, Kwon, Lee, Nam, Song, and Suh}]{kang2022korean}
Kang, T.; Kwon, E.; Lee, J.; Nam, Y.; Song, J.; and Suh, J. 2022.
\newblock Korean Online Hate Speech Dataset for Multilabel Classification: How Can Social Science Aid Developing Better Hate Speech Dataset?
\newblock arXiv:2204.03262.

\bibitem[{Kim et~al.(2021{\natexlab{a}})Kim, Guess, Nyhan, and Reifler}]{kim2021distorting}
Kim, J.~W.; Guess, A.; Nyhan, B.; and Reifler, J. 2021{\natexlab{a}}.
\newblock The distorting prism of social media: How self-selection and exposure to incivility fuel online comment toxicity.
\newblock \emph{Journal of Communication}, 71(6): 922--946.

\bibitem[{Kim(2022)}]{SmilegateAI2022KoreanUnSmileDataset}
Kim, S. 2022.
\newblock Korean UnSmile dataset: Human-annotated Multi-label Korean Hate Speech Dataset.
\newblock \url{https://github.com/smilegate-ai/korean_unsmile_dataset}.

\bibitem[{Kim et~al.(2021{\natexlab{b}})Kim, Shin, Sim, Jang, and Mingyoo}]{kpf_media_2021}
Kim, Y.; Shin, Y.; Sim, H.; Jang, Y.; and Mingyoo, P. 2021{\natexlab{b}}.
\newblock Media {Users} in {Korea} 2021.
\newblock Technical report, Korea Press Foundation.

\bibitem[{Korovkin, Park, and Kaganer(2023)}]{korovkin_towards_2023}
Korovkin, V.; Park, A.; and Kaganer, E. 2023.
\newblock Towards conceptualization and quantification of the digital divide.
\newblock \emph{Information, Communication \& Society}, 26(11): 2268--2303.

\bibitem[{Lee(2020)}]{lee2020kcbert}
Lee, J. 2020.
\newblock Kcbert: Korean comments bert.
\newblock In \emph{Annual Conference on Human and Language Technology}, 437--440. Human and Language Technology.

\bibitem[{Levenshtein(1966)}]{levenshtein1966binary}
Levenshtein, V.~I. 1966.
\newblock Binary codes capable of correcting deletions, insertions, and reversals.
\newblock \emph{Soviet Physics Doklady}, 10(8): 707--710.

\bibitem[{Masullo, Lu, and Fadnis(2021)}]{masullo2021does}
Masullo, G.~M.; Lu, S.; and Fadnis, D. 2021.
\newblock Does online incivility cancel out the spiral of silence? A moderated mediation model of willingness to speak out.
\newblock \emph{New Media \& Society}, 23(11): 3391--3414.

\bibitem[{Nielsen(2006)}]{nielsen_90-9-1_2006}
Nielsen, J. 2006.
\newblock The 90-9-1 Rule for Participation Inequality in Social Media and Online Communities.
\newblock \url{https://www.nngroup.com/articles/participation-inequality/}.
\newblock Accessed: 2024-01-06.

\bibitem[{Oh, Park, and Choi(2021)}]{oh2021digital}
Oh, S.-U.; Park, A.; and Choi, J. 2021.
\newblock Digital News Report in Korea 2021.
\newblock {\url{https://www.kpf.or.kr/front/research/selfDetail.do?seq=592216}}.

\bibitem[{Pak, André, and Beom(2021)}]{pak_digitalization_2021}
Pak, M.; André, C.; and Beom, J. 2021.
\newblock {DIGITALIZATION} {IN} {KOREA}: {A} {PATH} {TO} {BETTER} {SHARED} {PROSPERITY}?
\newblock Technical report, Korea Economic Institute of America.

\bibitem[{Papacharissi(2004)}]{papacharissi2004democracy}
Papacharissi, Z. 2004.
\newblock Democracy online: Civility, politeness, and the democratic potential of online political discussion groups.
\newblock \emph{New media \& society}, 6(2): 259--283.

\bibitem[{Rathje, Van~Bavel, and Van Der~Linden(2021)}]{rathje2021out}
Rathje, S.; Van~Bavel, J.~J.; and Van Der~Linden, S. 2021.
\newblock Out-group animosity drives engagement on social media.
\newblock \emph{Proceedings of the National Academy of Sciences}, 118(26): e2024292118.

\bibitem[{Rega, Marchetti, and Stanziano(2023)}]{rega2023incivility}
Rega, R.; Marchetti, R.; and Stanziano, A. 2023.
\newblock Incivility in online discussion: An examination of impolite and intolerant comments.
\newblock \emph{Social Media+ Society}, 9(2): 20563051231180638.

\bibitem[{Rossini(2022)}]{rossini_beyond_2022}
Rossini, P. 2022.
\newblock Beyond {Incivility}: {Understanding} {Patterns} of {Uncivil} and {Intolerant} {Discourse} in {Online} {Political} {Talk}.
\newblock \emph{Communication Research}, 49(3): 399--425.

\bibitem[{Rowe(2015)}]{rowe2015civility}
Rowe, I. 2015.
\newblock Civility 2.0: A comparative analysis of incivility in online political discussion.
\newblock \emph{Information, communication \& society}, 18(2): 121--138.

\bibitem[{Santana(2014)}]{santana2014virtuous}
Santana, A.~D. 2014.
\newblock Virtuous or vitriolic: The effect of anonymity on civility in online newspaper reader comment boards.
\newblock \emph{Journalism practice}, 8(1): 18--33.

\bibitem[{Scheerder, Van~Deursen, and Van~Dijk(2017)}]{scheerder2017determinants}
Scheerder, A.; Van~Deursen, A.; and Van~Dijk, J. 2017.
\newblock Determinants of Internet skills, uses and outcomes. A systematic review of the second-and third-level digital divide.
\newblock \emph{Telematics and informatics}, 34(8): 1607--1624.

\bibitem[{Valentino et~al.(2011)Valentino, Brader, Groenendyk, Gregorowicz, and Hutchings}]{valentino2011election}
Valentino, N.~A.; Brader, T.; Groenendyk, E.~W.; Gregorowicz, K.; and Hutchings, V.~L. 2011.
\newblock Election night’s alright for fighting: The role of emotions in political participation.
\newblock \emph{The journal of politics}, 73(1): 156--170.

\bibitem[{Van~Dijk(2006)}]{van_dijk_digital_2006}
Van~Dijk, J.~A. 2006.
\newblock Digital divide research, achievements and shortcomings.
\newblock \emph{Poetics}, 34(4-5): 221--235.
\newblock Publisher: Elsevier.

\bibitem[{Van~Mierlo(2014)}]{van_mierlo_1_2014}
Van~Mierlo, T. 2014.
\newblock The 1\% rule in four digital health social networks: an observational study.
\newblock \emph{Journal of medical Internet research}, 16(2): e2966.
\newblock Publisher: JMIR Publications Inc., Toronto, Canada.

\bibitem[{Yu, Wojcieszak, and Casas(2024)}]{yu2024partisanship}
Yu, X.; Wojcieszak, M.; and Casas, A. 2024.
\newblock Partisanship on social media: In-party love among American politicians, greater engagement with out-party hate among ordinary users.
\newblock \emph{Political Behavior}, 46(2): 799--824.

\end{thebibliography}

\appendix
\section{Appendix}
\subsection{Descriptive Statistics for the Comment Dataset}
\label{sec:appendix A1}
\subsubsection{Change in the Size of Comment Space}
\label{sec:appendix A1.1}
\begin{figure}[ht]
  \includegraphics[width=0.8\columnwidth]{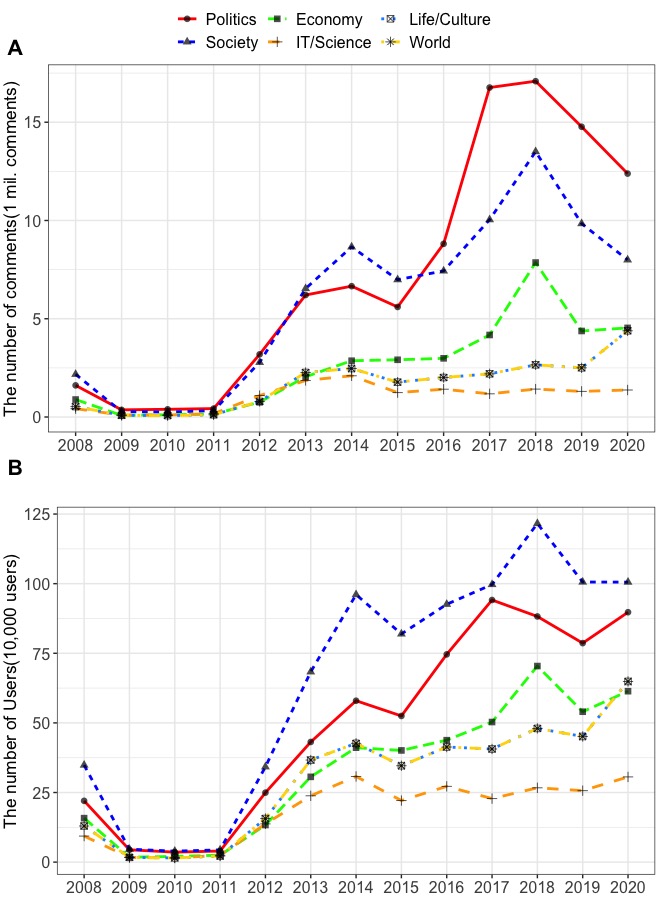}
  \centering
  \caption{Change in the size of comment space: A. Change in the number of comments over time. B. Change in the number of users over time}
  \label{fig9}
\end{figure}

The size of the comment space has grown rapidly over the years (Figure \ref{fig9}), and since our analysis focuses only on articles that received comments, we exclude users who did not engage in commenting. According to the widely cited 90-9-1 rule of online participation, approximately 90\% of users typically consume content without contributing. This suggests that our observed comment-based measures of participation inequality likely provide a lower-bound estimate.

\subsubsection*{Distribution of the Number of Comments}
\begin{figure}[ht]
  \includegraphics[width=0.9\columnwidth]{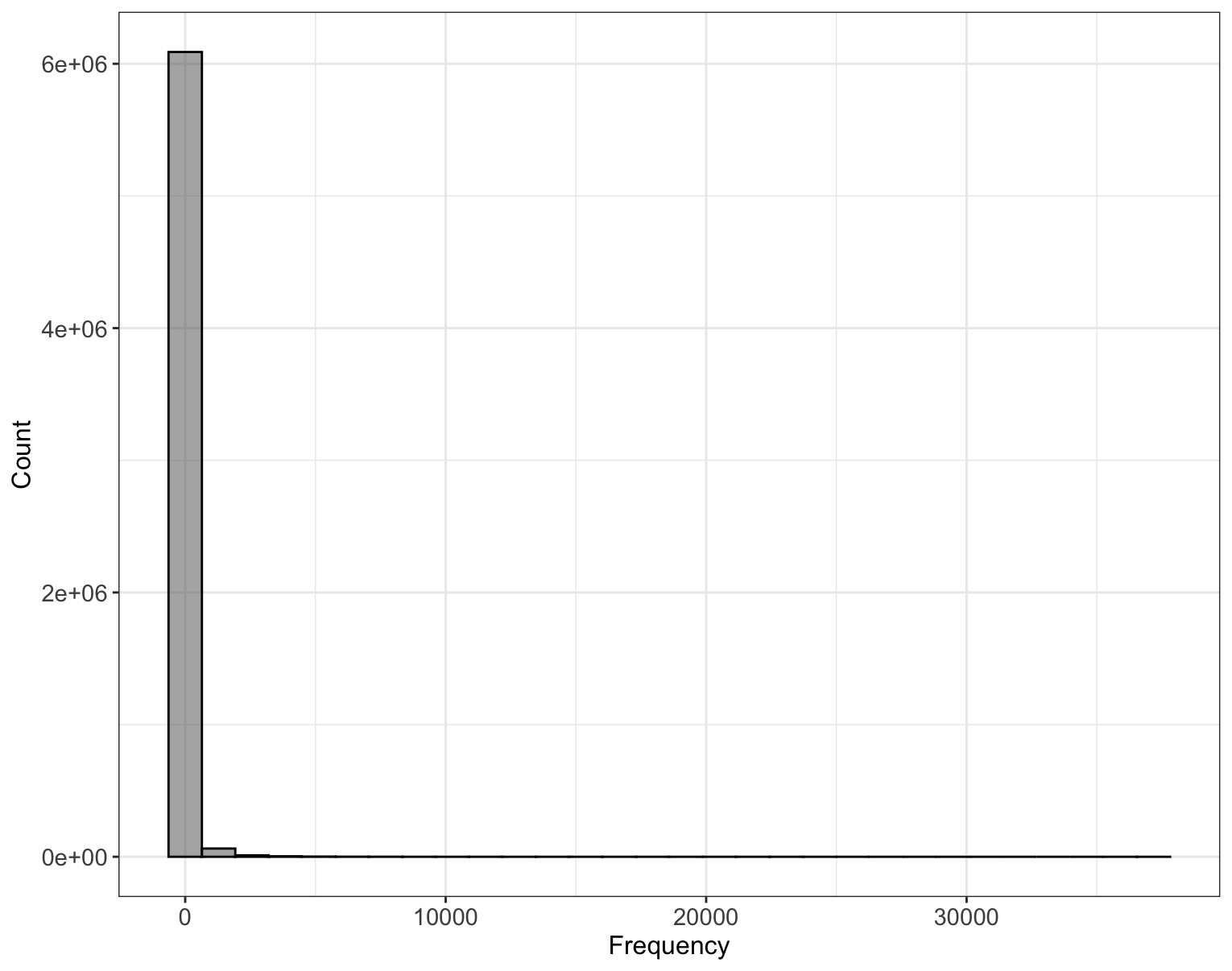}
  \centering
  \caption{Histogram of the Comment Frequency}
  \label{fig10}
\end{figure}

Online comment space is highly skewed. The histogram in Figure \ref{fig10} indicates that the majority of users post one or two comments. When dealing with a highly skewed distribution, it is generally more appropriate to consider specific percentiles, as there is a significant difference in values between the top and the bottom of the distribution. Hence, this paper compares only the top 10\% and the bottom 40\% groups.

\subsection{Check Bot-like Accounts}
Frequent commenters exhibit a range of behavioral patterns. While some highly active users contribute large volumes of comments across multiple news domains, others post large quantities of near-identical content repeatedly. To assess the extent of such anomalous behavior, we analyzed duplication and similarity scores among users in the frequent commenter group. 

For this analysis, we used the same stratified sample originally constructed for hostility classification. We focused on extremely active users (top 1\%), as bot-like accounts are more likely to be concentrated in this group. In contrast, users in the bottom 40\% group typically posted only once or twice, making the presence of bots in that group highly unlikely.

To measure the degree of similarity between user comments, we use the Levenshtein distance \citep{levenshtein1966binary}. This method calculates the minimum number of operations—such as deletion, insertion, and substitution—required to transform one string (\textit{string A}) into another (\textit{string B}). It provides a simple yet effective way to quantify textual similarity between pairs of comments. The distance is mathematically defined as follows:
\begin{equation}
\text{lev}(i, j) =
\begin{cases}
\max(i, j),\ \ \ \ \ \ \ \ \ \ \ \ \ \text{if  $min(i,j)=0$}\   \\
\min \left\{
\begin{array}{l}
\text{lev}(i - 1, j) + 1 \\
\text{lev}(i, j - 1) + 1 \\
\text{lev}(i - 1, j - 1) + \delta(a_i, b_j)
\end{array}
\right. & \text{o.w}
\end{cases}
\end{equation}
where
\begin{equation}
\delta(a_i, b_j) =
\begin{cases}
0 & \text{if } a_i = b_j, \\
1 & \text{otherwise}
\end{cases}
\end{equation}
Based on the number of duplicated comments and Levenshtein distance, we defined the rate of duplicated comments and the similarity score of user $i$ as follows:
\begin{equation}
\begin{aligned}
    Dup_i=1-\frac{No.\ Duplicated\ Comments_i}{No.\ Total\ Comments_i}\\
    \\
    \\
    \\
    Sim_i = 1 - \frac{\sum_{a_i,b_i\in c_i} \text{lev}(a_i, b_i)}{No.\ Total\ Comments_i}
\end{aligned}
\end{equation}
where $a_i$ and $b_i$ are distinct comments that user $i$ posted and $C_i$ is the collection of all comments that user $i$ posted.

\begin{figure}[H]
  \centering
  \begin{subfigure}{0.35\textwidth}
    \centering
    \includegraphics[width=\textwidth]{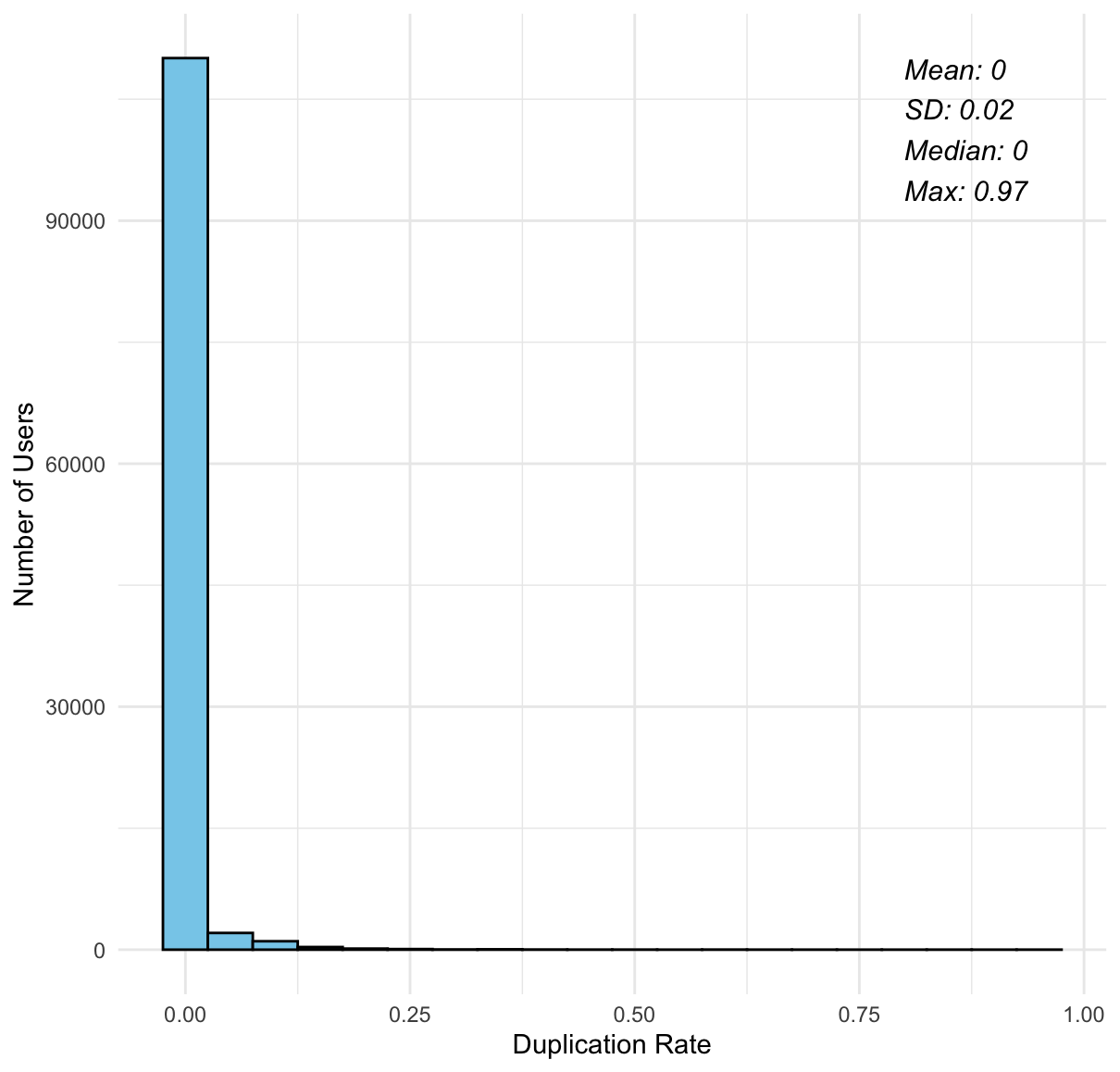}
    \caption{Duplication rate distribution among frequent users.}
    \label{fig:dup_rate}
  \end{subfigure}
  
  \vspace{1em}  

  \begin{subfigure}{0.44\textwidth}
    \centering
    \includegraphics[width=\textwidth]{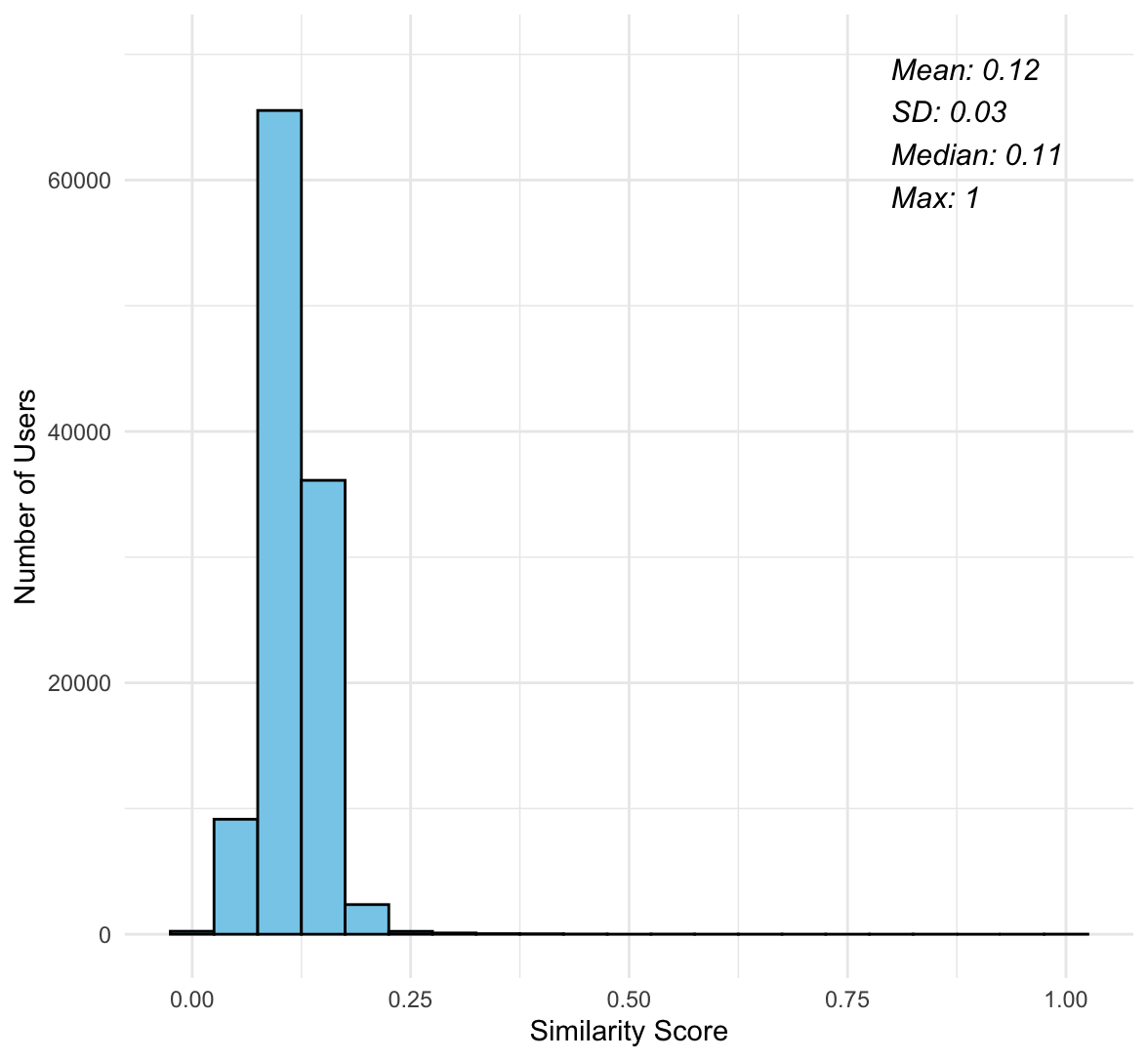}
    \caption{Similarity score distribution using Levenshtein distance.}
    \label{fig:sim_score}
  \end{subfigure}

  \caption{Exploratory analysis of user behavior: (a) exact duplication rates and (b) content similarity.}
  \label{fig:user_behavior}
\end{figure}

We find that, although a small subset exhibits highly repetitive behavior, the vast majority of users do not engage in abnormal posting patterns such as frequent comment duplication(Figure \ref{fig:user_behavior}). This suggests the presence of potentially bot-like accounts in the comment space. However, their prevalence appears to be limited and unlikely to significantly distort the overall patterns of participation or hostility observed in our analysis.

\begin{table*}[h!]
  \centering
  \setlength{\tabcolsep}{6pt}  
  
  \begin{subtable}[t]{\textwidth}
    \centering
    \small
    \begin{tabular}{@{}lccc|ccc|ccc|ccc@{}}
      \toprule
      \textbf{User Group} &
        \multicolumn{3}{c|}{\textbf{KC‑BERT Base}} &
        \multicolumn{3}{c|}{\textbf{KC‑BERT Large}} &
        \multicolumn{3}{c|}{\textbf{KoBERT}} &
        \multicolumn{3}{c}{\textbf{KoElectra}} \\
      \cmidrule(lr){2-4}\cmidrule(lr){5-7}\cmidrule(lr){8-10}\cmidrule(lr){11-13}
       & Civil & Uncivil & Hate & Civil & Uncivil & Hate & Civil & Uncivil & Hate & Civil & Uncivil & Hate \\
      \midrule
      Bottom 40\%       & 0.674 & 0.258 & 0.068  & 0.705 & 0.226 & 0.069  & 0.669 & 0.247 & 0.084  & 0.631 & 0.305 & 0.064  \\
      Top 1\% - Top 10\%        & 0.524 & 0.341 & 0.135  & 0.550 & 0.311 & 0.138  & 0.546 & 0.316 & 0.138  & 0.493 & 0.386 & 0.121  \\
      Top 1\%         & 0.486 & 0.337 & 0.177  & 0.506 & 0.303 & 0.191  & 0.511 & 0.318 & 0.171  & 0.455 & 0.389 & 0.156  \\
      \bottomrule
    \end{tabular}
    \caption{Raw proportions of comment categories by user group}
  \end{subtable}

  \vspace{1em}

  \begin{subtable}[t]{\textwidth}
    \centering
    \small
    \begin{tabular}{@{}lccc|ccc|ccc|ccc@{}}
      \toprule
      \textbf{Comparison} &
        \multicolumn{3}{c|}{\textbf{KC‑BERT Base}} &
        \multicolumn{3}{c|}{\textbf{KC‑BERT Large}} &
        \multicolumn{3}{c|}{\textbf{KoBERT}} &
        \multicolumn{3}{c}{\textbf{KoElectra}} \\
      \cmidrule(lr){2-4}\cmidrule(lr){5-7}\cmidrule(lr){8-10}\cmidrule(lr){11-13}
       & Civil & Uncivil & Hate & Civil & Uncivil & Hate & Civil & Uncivil & Hate & Civil & Uncivil & Hate \\
      \midrule
      Bottom 40\% vs Top 1\% - Top 10\%
       & 0.310 & 0.182 & 0.225  
       & 0.322 & 0.192 & 0.230  
       & 0.253 & 0.153 & 0.173  
       & 0.279 & 0.081 & 0.057  \\
      Bottom 40\% vs Top 1\%  
       & 0.383 & 0.173 & 0.341  
       & 0.410 & 0.175 & 0.373  
       & 0.323 & 0.158 & 0.265  
       & 0.355 & 0.177 & 0.301  \\
      \bottomrule
    \end{tabular}
    \caption{Cohen’s $h$ for civil, uncivil, and hateful categories}
  \end{subtable}

  \caption{Comparison of raw comment proportions and Cohen’s $h$ across models and user groups}
  \label{tab:app_combined}
\end{table*}

\subsection{Further Details about Hate Speech Dataset}
\subsubsection{Overall Structure of Dataset}
While the main text presents the distribution of individual hate speech labels, the dataset is broadly divided into two overarching categories: \textbf{Civil} and \textbf{Uncivil + Hate}. These two groups occur with nearly equal frequency, highlighting that the dataset is balanced in terms of overall tone. 

\subsubsection{Overlap among Uncivil and Hate Categories}
\begin{figure}[H]
  \includegraphics[width=\columnwidth]{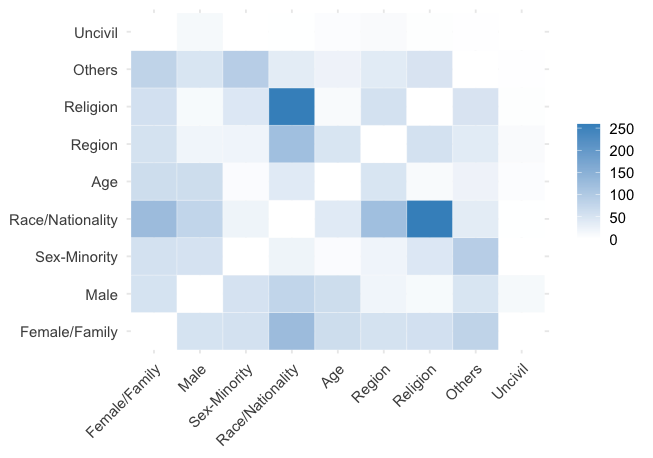}
  \caption{Co-occurrence Heatmap of Hate Speech Labels}
  \label{fig:cooc_heatmap}
  \caption*{\textit{Note.} The co-occurrence matrix is symmetric, as co-occurrence between labels A and B is identical to that between B and A.} 
\end{figure}

Given the multi-label nature of the dataset, a single comment can exhibit multiple forms of hostility. To better understand such overlaps, we visualize a label co-occurrence heatmap in Figure~\ref{fig:cooc_heatmap}, where each cell represents the number of comments assigned to both corresponding labels. 

Notably, \textit{religion} and \textit{race/nationality} often appear together, as do \textit{region} and \textit{race/nationality}. These patterns indicate the presence of overlapping hostile narratives targeting multiple social groups within individual comments. 

One thing to note is that while each comment in our dataset may be assigned multiple labels, and some overlapping patterns are observed, the average label cardinality is relatively low at 1.1, suggesting that most comments are tagged with only a single category.
\subsection{Robustness Check for Hate Speech Classification}
We conducted additional analyses using various pretrained language models to ensure our classification results are not model-specific. As shown in Table~\ref{tab:app_combined}, the core pattern persists across all models: the most active users consistently exhibit a higher proportion of hate comments than less active users. Although the magnitude of group differences varies slightly, as indicated by Cohen’s $h$, the direction and relative scale of the differences remain stable, supporting the robustness of our main findings
\subsection{Cohen's $h$}
Cohen’s $h$ \citep{cohen1988statistical} is a measure of effect size for differences between two proportions, defined as:
\[
h = 2 \cdot \arcsin(\sqrt{p_1}) - 2 \cdot \arcsin(\sqrt{p_2})
\]
The thresholds for interpreting $h$ are as follows:
\begin{itemize}
  \item $h = 0.20$: small effect size
  \item $h = 0.50$: medium effect size
  \item $h = 0.80$: large effect size
\end{itemize}

\end{document}